\documentclass[10pt]{iopart}
\pdfoutput=1
\usepackage[utf8]{inputenc}
\usepackage[T1]{fontenc}
\usepackage[american]{babel}
\usepackage{csquotes}

\usepackage{graphicx}
\usepackage{booktabs}
\usepackage{multirow}

\usepackage[usenames, dvipsnames, svgnames, table]{xcolor}
\usepackage{siunitx}
\usepackage[pdftex, breaklinks=true, linktocpage, colorlinks=true, urlcolor=blue, linkcolor=blue, citecolor=red]{hyperref}

\usepackage{dsfont}
\newcommand{\N}{\mathds{N}}

\hypersetup{
  pdftitle={Inception Neural Network for Complete Intersection Calabi-Yau 3-folds},
  pdfsubject={Deep Learning},
  pdfauthor={H. Erbin, R. Finotello},
  pdfkeywords={deep learning, Calabi-Yau manifold, string theory compactification, algebraic topology}
}

\begin{document}

\title[Inception Network for CICY]{Inception Neural Network for Complete Intersection Calabi-Yau 3-folds}

\author{H. Erbin, R. Finotello}
\address{Dipartimento di Fisica, Università di Torino and I.N.F.N.\ -- sezione di Torino, via P.\ Giuria 1, I-10125 Torino, Italy}
\ead{erbin@to.infn.it, riccardo.finotello@to.infn.it}

\begin{abstract}
We introduce a neural network inspired by Google's Inception model to compute the Hodge number $h^{1,1}$ of complete intersection Calabi-Yau (CICY) 3-folds.
This architecture improves largely the accuracy of the predictions over existing results, giving already $\SI{97}{\percent}$ of accuracy with just $\SI{30}{\percent}$ of the data for training.
Accuracy climbs to $\SI{99}{\percent}$ when using $\SI{80}{\percent}$ of the data for training.
This proves that neural networks are a valuable resource to study geometric aspects in both pure mathematics and string theory.
\end{abstract}

\section{Introduction}

The last few years witnessed the uprising of \emph{deep learning} as a very efficient method to elaborate, process and learn patterns in data~\cite{Goodfellow:2016:DeepLearning}.
While the underlying ideas behind neural networks are not recent~\cite{Rosenblatt:1958:Perceptron, Lecun:1998:LeNet}, larger databases, and computational capabilities together with new techniques led deep learning to pervade most fields of scientific research and industrial development.

Understanding geometrical structures is an emerging application of machine learning, referred to as \emph{geometric deep learning}~\cite{Bronstein:2017:GeometricDeepLearning, GeometricDeepLearning} when neural networks are used.
This is an important problem for different fields: for example in the industry (e.g.\ for $3d$ modelling of objects), computer science (e.g.\ for gradient optimisation~\cite{Lei:2020:GeometricUnderstandingDeep}), pure mathematics, and theoretical physics.
For this reason it is crucial to adapt existing techniques or to design new ones if needed.

In this paper, we focus on the computation of the Hodge number $h^{1,1}$ for complete intersection Calabi--Yau (CICY) $3$-folds~\cite{Green:1986:CICY}.
This is a challenging mathematical problem \emph{per se} because traditional methods from algebraic topology lead to complicated algorithms, without closed-form expressions in most cases.
Machine learning techniques give the possibility to speed up computations and to obtain hints to better understand the mathematical structures.
Moreover, Calabi--Yau manifolds, beyond being important mathematical objects, also have a distinguished role in string theory as they are needed to describe the compactified dimensions~\cite{Ibanez:2012:StringPhenomenology}.
In particular the general properties of the $4$-dimensional effective field theory are completely determined by the topology.
Given the complexity of the space of string vacua, developing faster and efficient computational techniques is essential in the search of the Standard Model (or an extension compatible with experiments) within string theory at low energy.
Finally, this type of objects is quite remote from typical data considered in machine learning, which calls for an evaluation of existing techniques in this context and, if they are not sufficient, the development of new approaches.

The CICY $3$-folds are appropriate for this task: since they have been completely classified~\cite{Candelas:1988:CICY, Green:1989:CICY, Anderson:2017:Fibrations}, they provide a simple playground where it is possible to test different machine learning techniques.
The goal of this paper is to continue the study started in~\cite{He:2017:Landscape, Bull:2018:CICY3Folds}, which used machine learning techniques to compute $h^{1,1}$ (see also~\cite{Bull:2019:GettingCICYHigh, He:2019:DistinguishingEllipticFibrations, Krippendorf:2020:DetectingSymmetriesNeural} for other papers on CICY $3$-folds).
Related applications on the study of cohomology groups are~\cite{Ruehle:2017:EvolvingNeuralNetworks, Klaewer:2019:MachineLearningLine, Brodie:2020:MachineLearningLine}.
For an introduction to machine learning and its applications to string theory, we refer to the excellent review~\cite{Ruehle:2020:DataScienceApplications}.

\begin{figure}[tbp]
\centering
\includegraphics[width=0.7\linewidth]{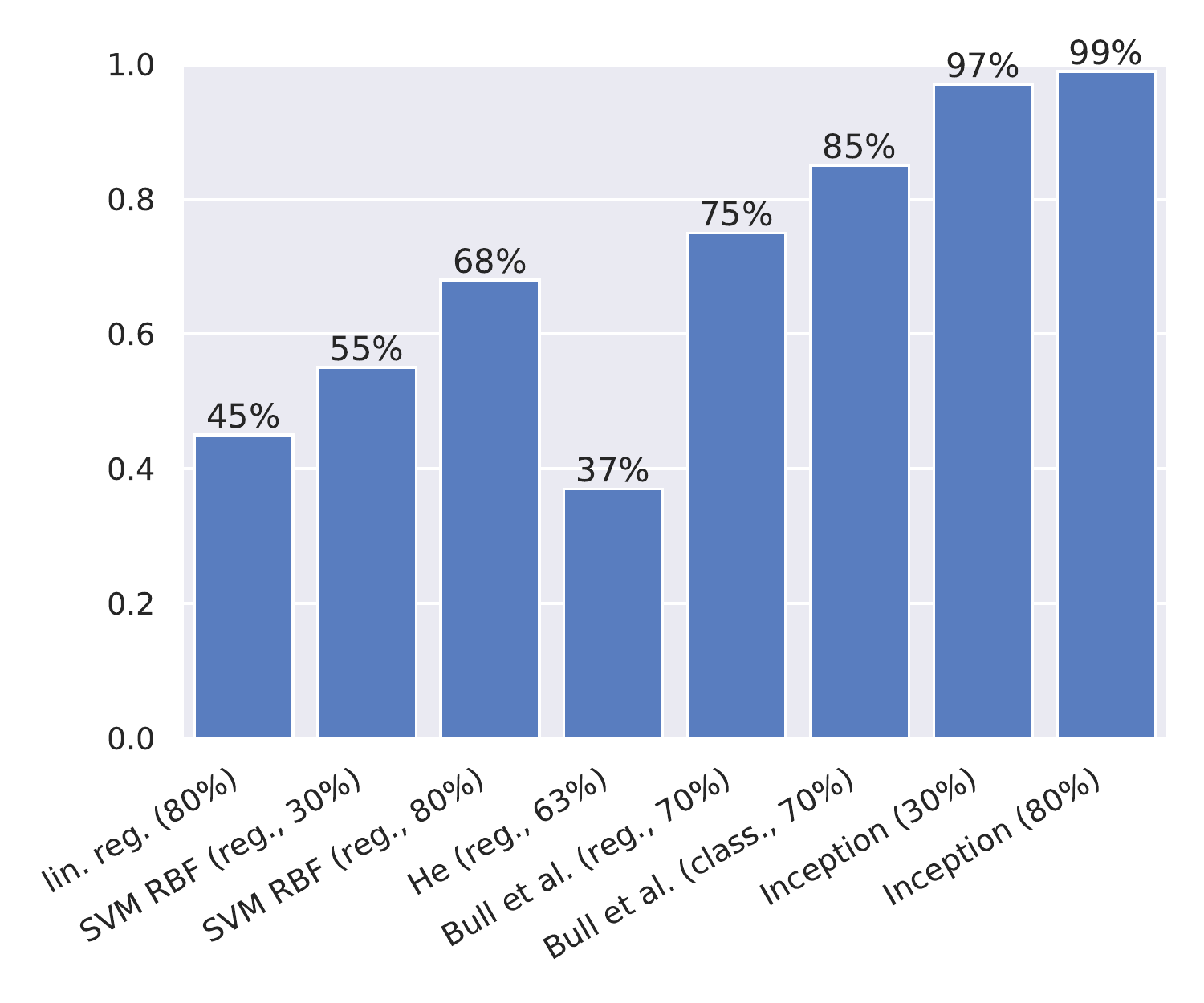}
\caption{%
  Accuracy reached by different models.
  The percentage in parenthesis indicates the ratio of training data.
  ``He'' refers to~\cite{He:2017:Landscape}, ``Bull et al.''\ to~\cite{Bull:2018:CICY3Folds}.
  Each model except the Inception one keeps outliers in the training set (the effects is marginal in linear regression and SVM).
}
\label{fig:intro:compare}
\end{figure}

Most breakthroughs in AI and industrial applications of deep learning usually followed the discovery of a new network model.
This is particularly true in computer vision where convolutional, Inception and residual networks~\cite{Lecun:1998:LeNet, Szegedy:2014:InceptionNetwork, Szegedy:2015:RethinkingInceptionArchitecture, Szegedy:2017:Inceptionv4InceptionResNetImpact, He:2016:ResNet} have been major cornerstones.
In this work, we introduce an alternative version of Google's \emph{Inception} network~\cite{Szegedy:2014:InceptionNetwork, Szegedy:2015:RethinkingInceptionArchitecture, Szegedy:2017:Inceptionv4InceptionResNetImpact} (see~\cite{Ruehle:2020:DataScienceApplications} for a review) to predict $h^{1,1}$ from the configuration matrix of CICY 3-folds.
Using $\SI{30}{\percent}$ of training data, we reach close to $\SI{97}{\percent}$ accuracy on the predictions, improving by a large measure previous results~\cite{He:2017:Landscape, Bull:2018:CICY3Folds} with much less training data and parameters ($\num{\approx 234000}$).
Using $\SI{80}{\percent}$ for the data for training we obtain $\SI{99}{\percent}$ accuracy.

This must be compared with the following accuracies:
$\SI{37}{\percent}$ (regression, fully connected network, $\num{\approx 280000}$ parameters, $\SI{63}{\percent}$ training data) in~\cite{He:2017:Landscape},
$\SI{75}{\percent}$ (regression, fully connected network, $\num{\approx 1580000}$ parameters, $\SI{70}{\percent}$ training data) and $\SI{85}{\percent}$ (classification, convolutional network, $\SI{70}{\percent}$ training data) in~\cite{Bull:2018:CICY3Folds} (Figure~\ref{fig:intro:compare}).
More generally, we found that the Inception-like network performs much better than any other machine learning algorithm, even after feature engineering~\cite{Erbin:2020:MachineLearningMethodological}: the best algorithm after neural networks is SVM with an RBF kernel, which reaches $\SI{68}{\percent}$ accuracy with $\SI{80}{\percent}$ of training data~\cite{Bull:2018:CICY3Folds, Erbin:2020:MachineLearningMethodological}.
This shows that neural networks are able to make accurate predictions for Hodge numbers, as long as the correct architecture is found.
This opens the door to new applications to theoretical physics and mathematics which may lead to even further progress.

The code is written in Python and relies on the following packages: \texttt{scikit-learn}~\cite{Pedregosa:2011:ScikitLearn}, \texttt{tensorflow}~\cite{Google:2015:Tensorflow} (and its high level API, \texttt{keras}~\cite{Package:Keras}) and the \texttt{scipy} ecosystem for visualisation and computations~\cite{Scipy:2020:Scipy}.

\section{General Setup}

The dataset~\cite{Candelas:1988:CICY, Green:1989:CICY} is made of 7890 CICY 3-folds, described by their configuration matrices and their topological properties, including the Hodge numbers $h^{1,1}$ and $h^{2,1}$.
We focus on predicting the Hodge number $h^{1,1} \in \N$, which lies in the closed interval $[0,~19]$ with $18$ distinct values (with $h^{1,1} = 17, 18$ not present), from the configuration matrix:
\begin{equation}
	\left[
		\begin{array}{c|ccc}
		\mathds{P}^{n_1} & a_1^1 & \cdots & a_k^1
		\\
		\vdots & \vdots & \ddots & \vdots
		\\
		\mathds{P}^{n_m} & a_1^m & \cdots & a_k^m
		\end{array}
	\right],
	\quad
	a^r_\alpha \in \N
	\quad \longrightarrow \quad
	h^{1,1} \in \N.
\end{equation}
The configuration matrix describes the CICY as the intersection of $k$ hypersurfaces, characterised by a system of homogeneous polynomial equations, inside the ambient space $\mathds{P}^{n_1} \times \cdots \times \mathds{P}^{n_m}$, where $m$ denotes the number of complex projective spaces.
The coefficients $a^r_\alpha$ of the matrix denote the power of the coordinates of each projective space entering each polynomial equation.
This data is sufficient to characterise the topology.
For more information on CICY we refer the reader to the literature~\cite{Green:1987:CalabiYauManifoldsComplete, Green:1987:PolynomialDeformationsCohomology, Candelas:1988:CICY, Green:1989:CICY, Anderson:2017:FibrationsCICYThreefolds, Anderson:2018:TASILecturesGeometric}.

We consider the problem as a regression task and not as a classification task even if the outputs are integers.
Indeed, the latter requires knowledge of all possible Hodge numbers which can appear and prevents any extrapolation, which is not desirable in the current context.
Since regression algorithms output a real number, it is necessary to map predictions to integers before comparing with the real values.

The dataset is split into three subsets: one for training (used to learn the optimal model weights with gradient descent), one for validation (hyperparameter tuning and early stopping in neural networks), and one for testing.

In the following section, we discuss a few properties of the dataset which play an important role in the training of the neural network introduced in the next section.

\subsection{Exploratory Data Analysis}

\begin{figure}[tbp]
\centering
\includegraphics[width=0.6\linewidth]{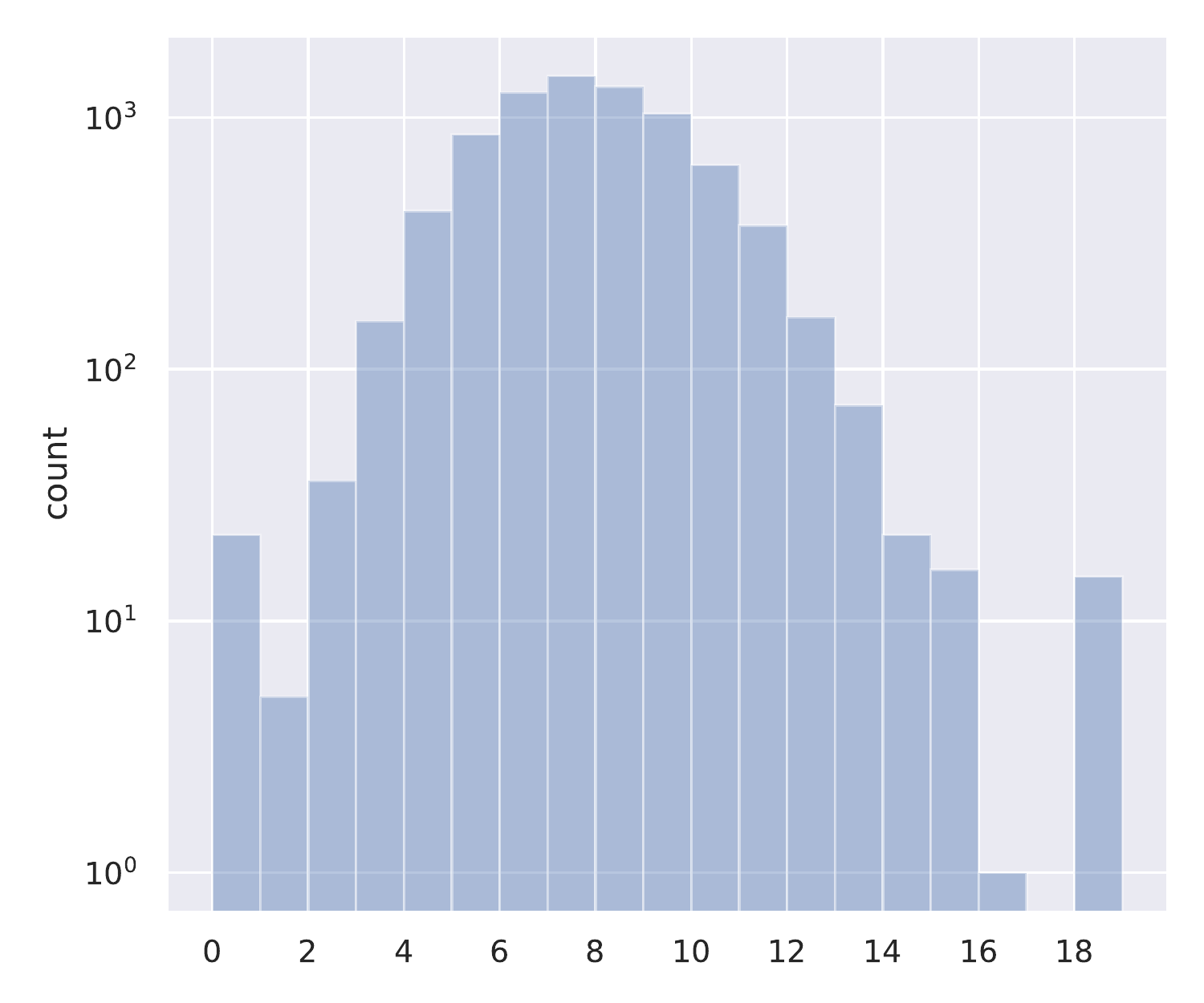}
\caption{%
  Distribution of $h^{1,1}$ (log scale).
}
\label{fig:eda:dist}

\end{figure}

\begin{figure}[tbp]
\centering
\includegraphics[width=0.6\linewidth, trim={0 0 6in 0}, clip]{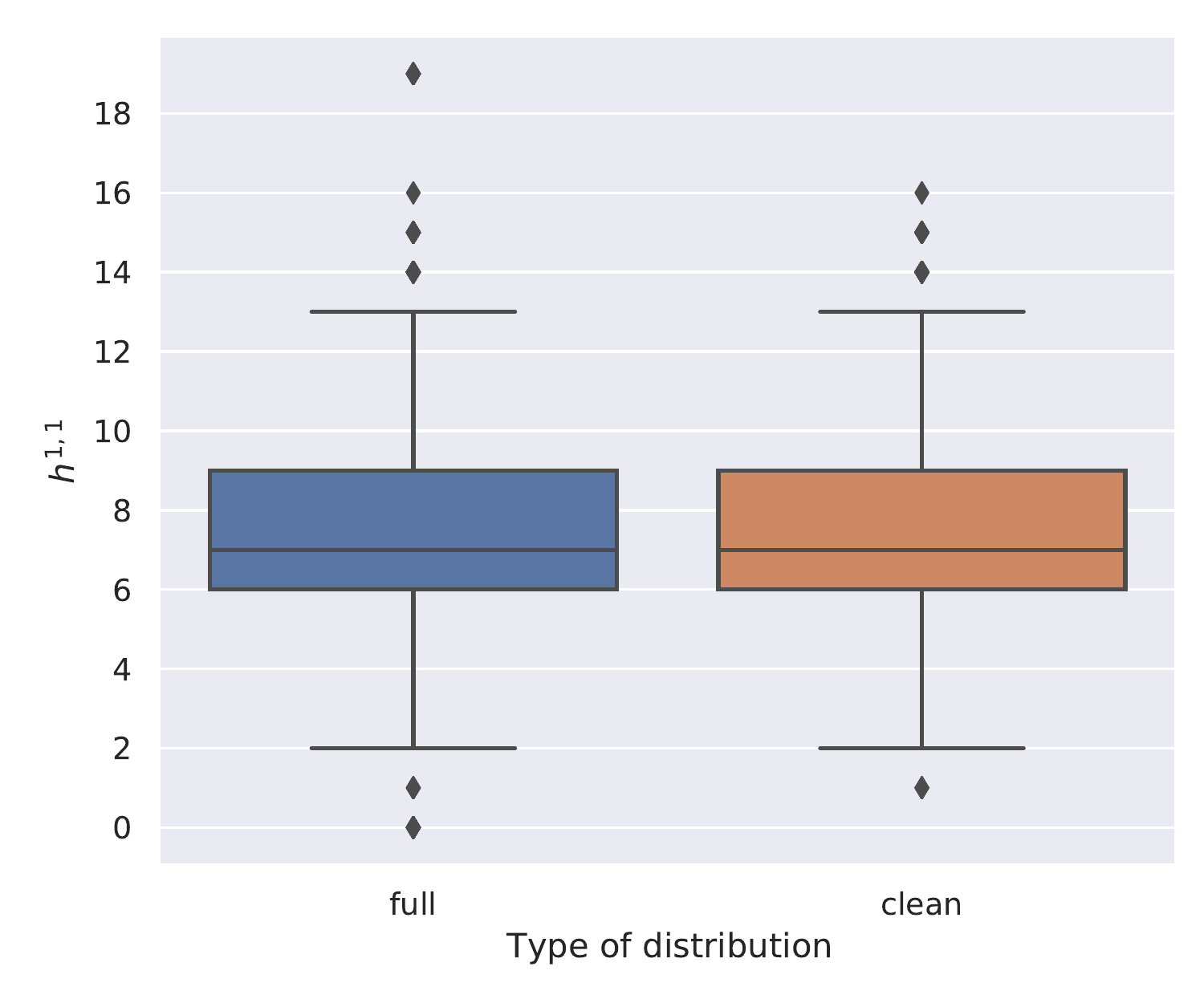}
\caption{%
  Whisker plot for the distribution of $h^{1,1}$ before (\emph{full}) and after (\emph{clean}) removing the outliers.
  The coloured boxes highlight the interval between the first and third quartiles while the internal horizontal line represents the median value.
  The ``whiskers'' delimit the interquartile range, while isolated points mark the remaining outliers.
}
\label{fig:eda:labels}
\end{figure}

The first step before writing the neural network is to better understand the data.
Displaying the distribution of the Hodge numbers (Figure~\ref{fig:eda:dist}) and the whisker plot in the left side of Figure~\ref{fig:eda:labels}, one finds the presence of outliers at small and high Hodge numbers.
Outliers can strongly impede the learning process of most algorithms and they must be handled with care.
In this paper we obtained the best accuracy by simply removing them from the training data (but keeping them in the test set).

The outliers fall into two classes.
First, the product spaces are recognisable by having vanishing Hodge numbers $h^{1,1} = h^{2,1} = 0$ and a block-diagonal configuration matrix.\footnote{Note that $h^{1,1} = 0$ is not the the actual value of $h^{1,1}$ but indicates merely that the CICY is factorizable into products of tori and K3 surfaces.}
Second we deal with manifolds with high Hodge numbers.
We keep only manifolds such that $h^{1,1} \in [1, 16]$ and $h^{2,1} \in [15, 86]$ in the training data.
Over the full dataset only $39$ samples are excluded, or $\SI{0.49}{\percent}$.
Hence training samples are taken as a subset of the distribution given in the right side of Figure~\ref{fig:eda:labels}.
We expect systematical errors on test samples among outliers but they are too few to drastically impact the accuracy.


\subsection{Baseline}

It is important to design a simple baseline model to quantify the gain of using a neural network.
Here, we consider a linear regression with $\ell_1$ regularisation with parameter $\num{2e-4}$ and without intercept.
Integers are obtained by flooring the predictions to the next lower integers.
We obtain $\SI{47}{\percent}$ to $\SI{51}{\percent}$ accuracy using $\SI{20}{\percent}$ to $\SI{80}{\percent}$ of the data for training.

\begin{figure}[tbp]
  \centering
  \includegraphics[width=0.7\linewidth]{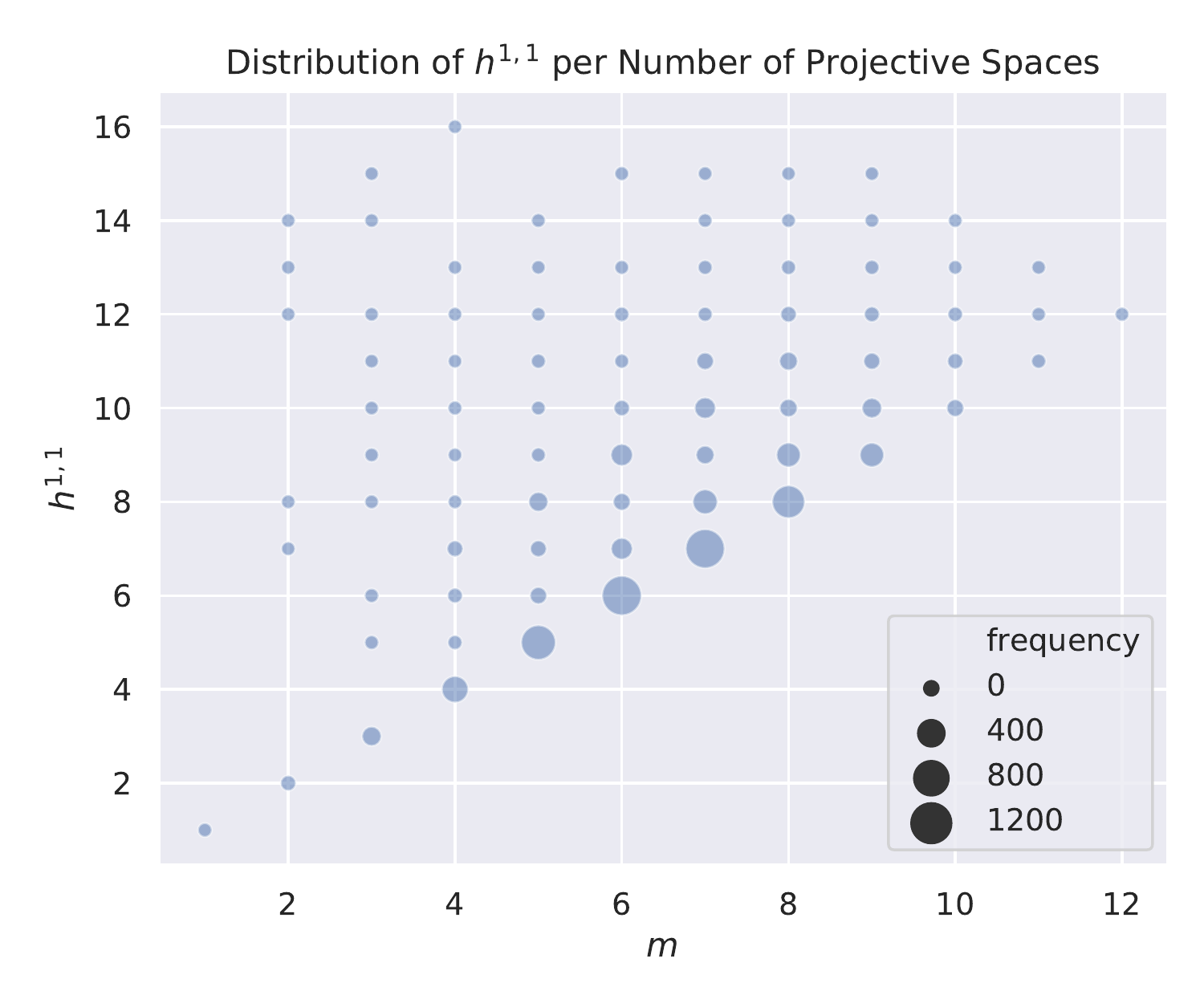}
  \caption{%
    Scatter plots of $h^{1,1}$ versus $m$.
    Manifolds on the diagonal are favourable.
  }
  \label{fig:eda:scatter}
\end{figure}

Moreover a simple analysis~\cite{Erbin:2020:MachineLearningMethodological} shows that the number of projective spaces $m$ (number of rows of the matrix) is an important feature.
Performing a linear regression with $\ell_1$ weight of $\num{1.0}$, we obtain $\SI{63}{\percent}$ of accuracy.
This is related to a known mathematical result~\cite{Anderson:2017:FibrationsCICYThreefolds} stating that the so-called favourable matrices have $h^{1,1} = m$ (in the dataset from~\cite{Candelas:1988:CICY, Green:1989:CICY}, there are $4874$ favourable matrices, visible on the diagonal of the scatter plot in Figure~\ref{fig:eda:scatter}).
If it had not been known, the linear regression could have led to conjecture that this formula -- indeed, conjecture generation is another distinguished use of machine learning techniques for theoretical physics~\cite{Carifio:2017:MachineLearningString, Brodie:2020:MachineLearningLine}.
Note that SVM with RBF kernel is the best ML algorithm outside neural networks but improves only marginally over linear regression (Figure~\ref{fig:intro:compare})~\cite{Erbin:2020:MachineLearningMethodological}.

\section{Inception Neural Network}
\label{sec:inc-nn}

In this section, we introduce a new deep learning architecture capable of predicting accurately $h^{1,1}$ from the configuration matrix of the CICY manifolds.
Though different both in purpose and in definition, the model is inspired by Google's \emph{Inception Network}~\cite{Szegedy:2014:InceptionNetwork, Szegedy:2015:RethinkingInceptionArchitecture, Szegedy:2017:Inceptionv4InceptionResNetImpact}.
This deep neural network uses inception modules performing different concurrent convolutional operations to enhance, process, and rearrange its input (in Google's case, images to be classified over 1000 classes in the \emph{ImageNet} repository).
This architecture encountered great success as it obtained results much better than any other machine learning algorithm until then.
Modifications of the original model brought even higher accuracy and enhancement of computer vision capabilities.
We refer the reader to~\cite{Ruehle:2020:DataScienceApplications} for a review of Inception networks.

We arrived at this network by going through neural network architectures used in computer vision.
Indeed, the configuration matrix being a matrix of integers, it resembles an image with one channel.
Since a sequential convolutional network does not reach a sufficient accuracy and needs a lot of training data, the Inception is the next natural step.
Its structure has been guided by the form of the configuration matrix (see subsection~\ref{sec:ablation} for more details).

Adapting this network to our problem, we obtain close to $\SI{100}{\percent}$ accuracy already by training with only $\SI{30}{\percent}$ of the data, which is much higher than existing results~\cite{He:2017:Landscape, Bull:2018:CICY3Folds}.
A more general machine learning analysis of this problem will appear in~\cite{Erbin:2020:MachineLearningMethodological}.

\subsection{Architecture}
\label{sec:architecture}

\begin{figure}[tbp]
\centering
\includegraphics[width=\linewidth]{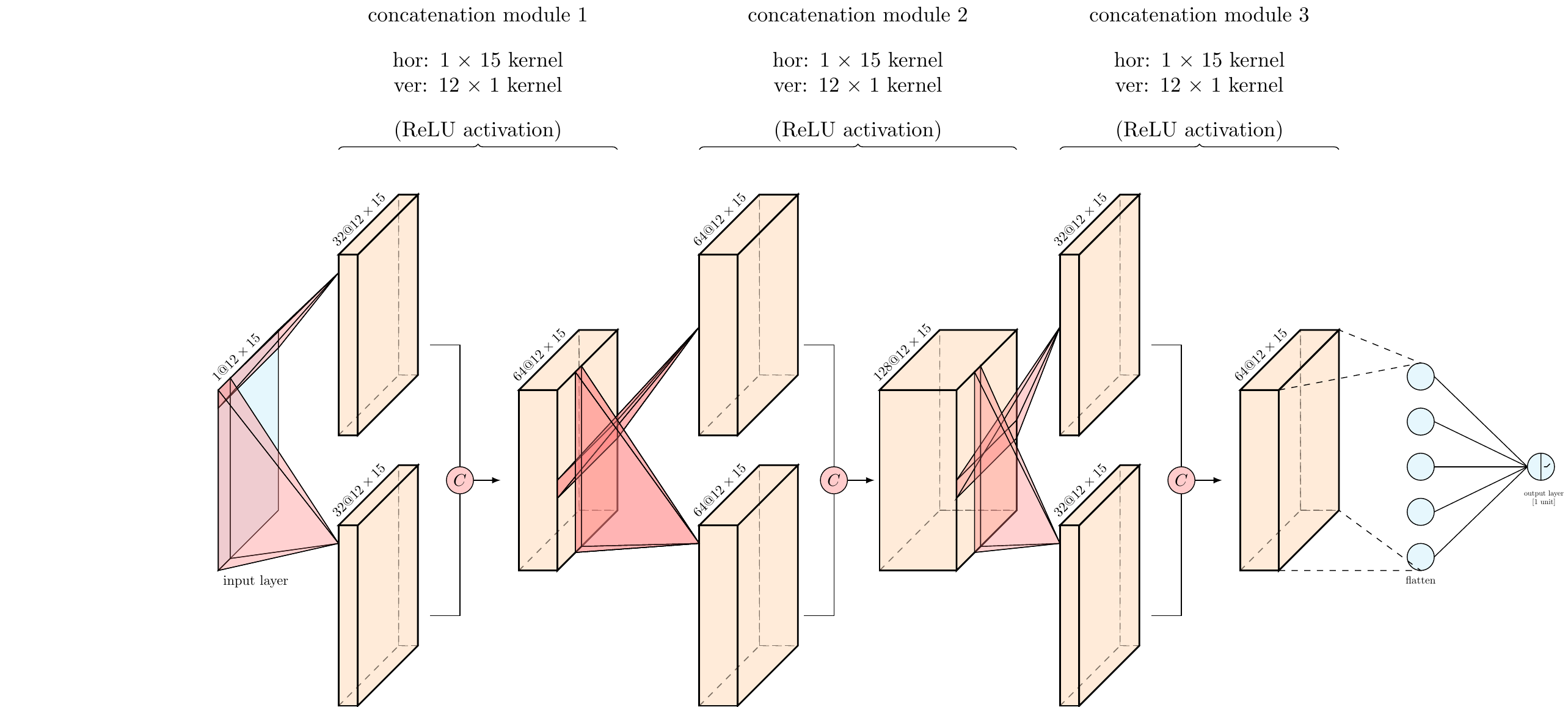}
\caption{Schematic figure of the inception model used to predict $h^{1,1}$ from configuration matrices.}
\label{fig:cnn:inception_img}
\end{figure}

The architecture is schematically depicted in Figure~\ref{fig:cnn:inception_img}: it is divided into three \emph{inception} modules followed by an output layer with a single unit for the prediction of the Hodge number.

The first layer takes the configuration matrices as input, which are represented as tensors of shape $(12, 15, 1)$ (matrices with a single channel).
Next two parallel convolutions (shown in red in Figure~\ref{fig:cnn:inception_img}) are performed: one over the rows ($12 \times 1$ kernel, processing each projective space at a time) and one over the columns ($1 \times 15$ kernel, processing each equation of the polynomial system at a time).
The outputs of both layers are concatenated together over the channel dimension.
These two steps form an inception module, which is repeated $3$ times in total, with respectively $32$, $64$, and $32$ filters.
All convolutional layers and the final layer are followed by a ReLU activation function and each concatenation by a batch normalisation with momentum $0.99$.
A dropout layer with a rate $0.2$ after the last inception module and before flattening the results, to connect it to the final output layer.
Finally, all layers have $\ell_1$ and $\ell_2$ regularisation, respectively with weights \num{e-4} and \num{e-3}.

Table~\ref{tab:cnn:inception_summary} summarises the network and the number of parameters in each layer.
The network has $\num{\approx 234000}$ parameters, which is less than previous proposals~\cite{He:2017:Landscape, Bull:2018:CICY3Folds}.
This is achieved by using only convolutional layers with relatively small kernels.

\begin{table}[tbp]

  \centering
    \begin{tabular}{@{}llll@{}}
      \toprule
      & \textbf{layer} & \textbf{shape} & \textbf{parameters} \\ \midrule\midrule
       \emph{input}                     & Input          & (12, 15, 1)    & \num{0}             \\ \midrule
       \multirow{4}{*}{\emph{module 1}} & Conv2D@12x1    & (12, 15, 32)   & \num{416}           \\
      & Conv2D@1x15    & (12, 15, 32)   & \num{512}           \\
      & concatenate    & (12, 15, 64)   & \num{0}             \\
      & BatchNorm      & (12, 15, 64)   & \num{256}           \\ \midrule
       \multirow{4}{*}{\emph{module 2}} & Conv2D@12x1    & (12, 15, 64)   & \num{49216}         \\
      & Conv2D@1x15    & (12, 15, 64)   & \num{61504}         \\
      & concatenate    & (12, 15, 128)  & \num{0}             \\
      & BatchNorm      & (12, 15, 128)  & \num{512}           \\ \midrule
       \multirow{4}{*}{\emph{module 3}} & Conv2D@12x1    & (12, 15, 32)   & \num{49184}         \\
      & Conv2D@1x15    & (12, 15, 32)   & \num{61472}         \\
      & concatenate    & (12, 15, 64)   & \num{0}             \\
      & BatchNorm      & (12, 15, 64)   & \num{256}           \\ \midrule
       \emph{dropout}                   & Dropout        & (12, 15, 64)   & \num{0}             \\ \midrule
       \multirow{2}{*}
       {\emph{fully connected}}         & Flatten        & (11520,)       & \num{0}             \\
      & Dense          & (1,)           & \num{11521}         \\ \midrule\midrule
       \emph{total parameters}          &                &                & \num{234849}        \\
       \emph{trainable parameters}      &                &                & \num{234337}        \\
       \emph{non trainable parameters}  &                &                & \num{512}           \\ \bottomrule
    \end{tabular}
  \caption{%
     Summary of the model with the number of parameters for each layer.
  }
  \label{tab:cnn:inception_summary}
\end{table}

Note that there are no pooling layers.
Convolutions use \texttt{same} as padding value which allows us to keep the same size $(12, 15)$ as the input.
The output layer is followed by a ReLU activation function which forces the result to be positive, as it should be for Hodge numbers.

This architecture has two evident advantages over a fully connected (FC) network or even a more classical convolutional structure.
First, the network concurrently learns different representations and automatically combines them in more complex representations.
Second, the number of parameters is extremely restricted.

\subsection{Training and validation strategy}

We use a \emph{holdout} validation strategy: the dataset is divided into three subsets for training (gradient descent to optimize the neural network's weights), validation (early stopping, and hyperparameter tuning) and testing purposes (final assessment of our model).
We retain respectively $\SI{80}{\percent}$ of all samples for training, $\SI{10}{\percent}$ for validation, and $\SI{10}{\percent}$ for testing.

Before feeding the configuration matrix to the neural network, we first remove the outliers as discussed previously.
We have tried to rescale the matrix by dividing by the highest entry ($5$), but this does not bring any significant improvement.

We did not use any data augmentation.
Adding matrices with permutations of rows and columns seem to decrease the performance of the neural network: one possible explanation is that matrix components are ordered lexicographically~\cite{Candelas:1988:CICY}.
Moreover, we did not generated more matrices using mathematical equivalences~\cite{Candelas:1988:CICY} since the final accuracy is high enough.

Hyperparameter tuning (number of inception modules and filters, dropout rate, etc.) has been performed by hand by evaluating several models on the validation set.
After finding the appropriate architecture, described in the previous subsection, we have also evaluated the accuracy by training with $\SI{30}{\percent}$ and $\SI{50}{\percent}$ of the data (keeping always $\SI{10}{\percent}$ for the validation set, necessary for early stopping).

\begin{figure}[tbp]
\centering
\includegraphics[width=0.475\linewidth]{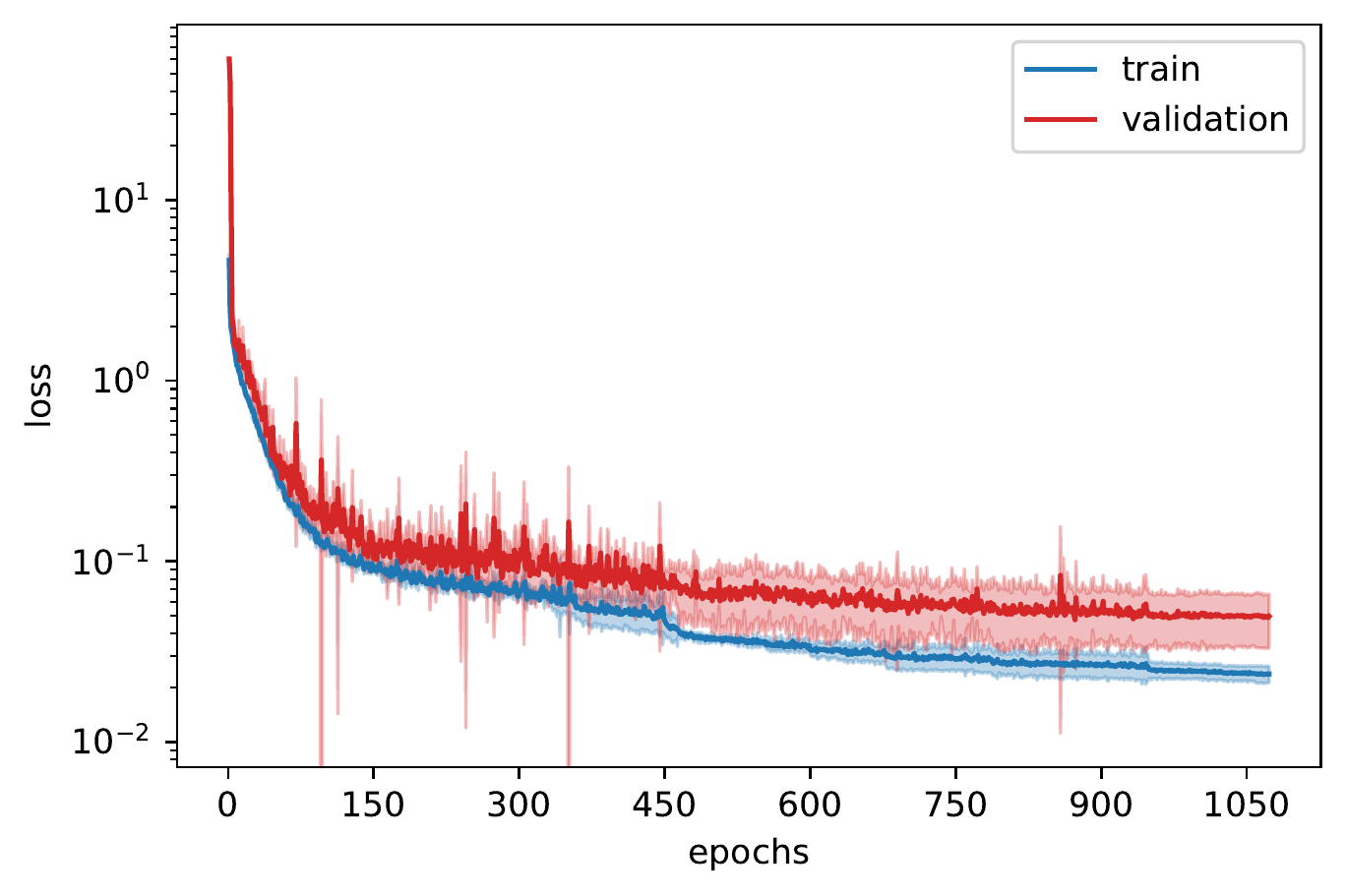}
\hfill
\includegraphics[width=0.475\linewidth]{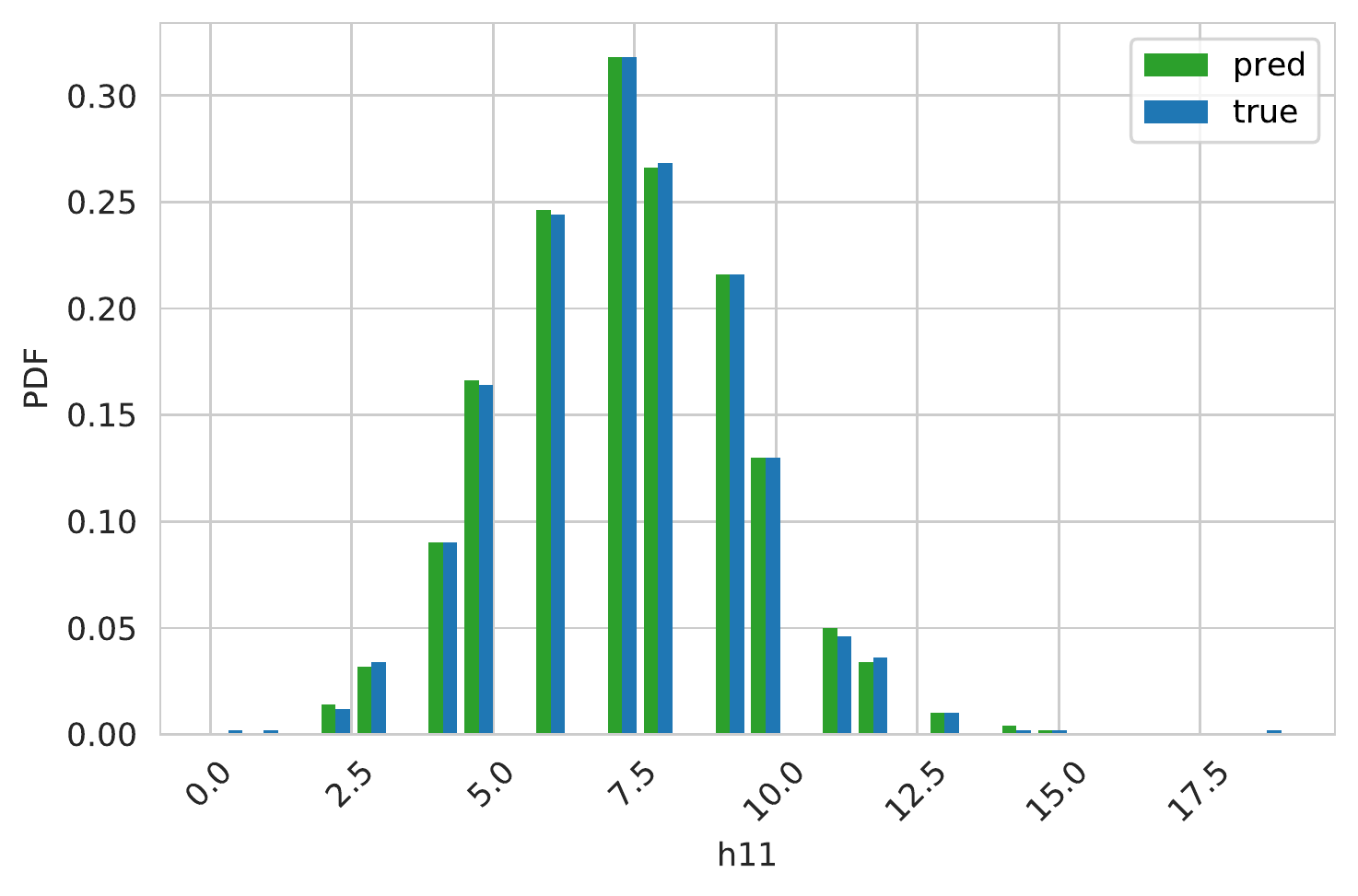}
\caption{%
  On the left we show the evolution of the loss evaluated on the training (blue curve) and evaluation (red curve) sets during training with $\SI{80}{\percent}$ training data (the colored area denotes the $1\sigma$ region).
  On the right we present the distribution of the real and predicted values for the Inception network ($\SI{80}{\percent}$ training data).
}
\label{fig:inc:loss}
\end{figure}

The neural network is trained using the \emph{Adam}~\cite{Kingma:2014:Adam} optimiser with default parameters, initial learning rate \num{e-3} and a batch size of \num{32}.
We use the mean squared error of the predictions as a loss function.
The learning rate is reduced by a factor of $0.3$ when the validation loss does not decrease during $75$ epochs.
We also use early stopping: the network is trained until the validation loss does not decrease for $200$ epochs, restoring the weights associated with the lowest validation loss.

Predictions are obtained by averaging the results of $5$ neural networks (bagging), which allows us to reduce the variance and obtain the standard deviation of the results.
Since predictions are real numbers at this point, they are rounded to the closest integers before comparing them with the real value.
The performance of the model is measured by the accuracy, which is the ratio of predictions matching exactly the real values.

Finally, we will also provide learning curves for the neural network described in the previous section.
For this, we split the dataset into training and validation subsets with different relative ratios and we compute the accuracy on both sets after training.
In each case, we keep $\SI{10}{\percent}$ of the training data for early stopping.
Exception made for this, the rest of the setup is the same.

\subsection{Results}

In the left plot in Figure~\ref{fig:inc:loss}, we show the evolution of the training and validation loss (mean squared error) during training.
Curiously the mean absolute error is smaller for the validation set.


\begin{table}[tbp]
\centering
\begin{tabular}{@{}c|ccc@{}}
\toprule
\emph{training data} & \textbf{Fully connected}  & \textbf{Convolution} & \textbf{Inception}  \\ \midrule
$\SI{80}{\percent}$    & $\SI{\approx 77}{\percent}$ & $\SI{92.5}{\percent}$  & $\SI{98.7}{\percent}$ \\
$\SI{50}{\percent}$    & $\SI{\approx 74}{\percent}$ & $\SI{84.9}{\percent}$  & $\SI{98.3}{\percent}$ \\
$\SI{30}{\percent}$    & $\SI{\approx 68}{\percent}$ & $\SI{78.5}{\percent}$  & $\SI{97.6}{\percent}$ \\ \bottomrule
\end{tabular}
\caption{%
  Accuracy for the Inception neural network for different sizes of the training dataset, with
  standard deviations between $\SI{0.1}{\percent}$ to $\SI{0.5}{\percent}$.
  Results obtained for other models are added for comparison: fully connected network~\cite{Bull:2018:CICY3Folds} (read from Figure~1), convolutional network~\cite{Erbin:2020:MachineLearningMethodological}.
  See also Figure~\ref{fig:intro:compare}.
}
\label{tab:cnn:inception_results}
\end{table}

The agreement between the predictions and real values is excellent on the test fold.
The distributions are displayed on the right of Figure~\ref{fig:inc:loss}.
The results at different ratios of training data are given in Table~\ref{tab:cnn:inception_results}, where we also display the accuracy for other regression models: the fully connected network from~\cite{Bull:2018:CICY3Folds} and an improved sequential convolutional network described in~\cite{Erbin:2020:MachineLearningMethodological} (see also the introduction for more details).
Even though the sequential model can already achieve very high accuracy, the Inception network performs even better with fewer parameters and much less training data.
The learning curve is given in Figure~\ref{fig:cnn:lc}: it does not show signs of overfitting and clearly demonstrates the quick convergence to almost $\SI{100}{\percent}$ accuracy.

\begin{figure}[tbp]
\centering
\includegraphics[width=\linewidth]{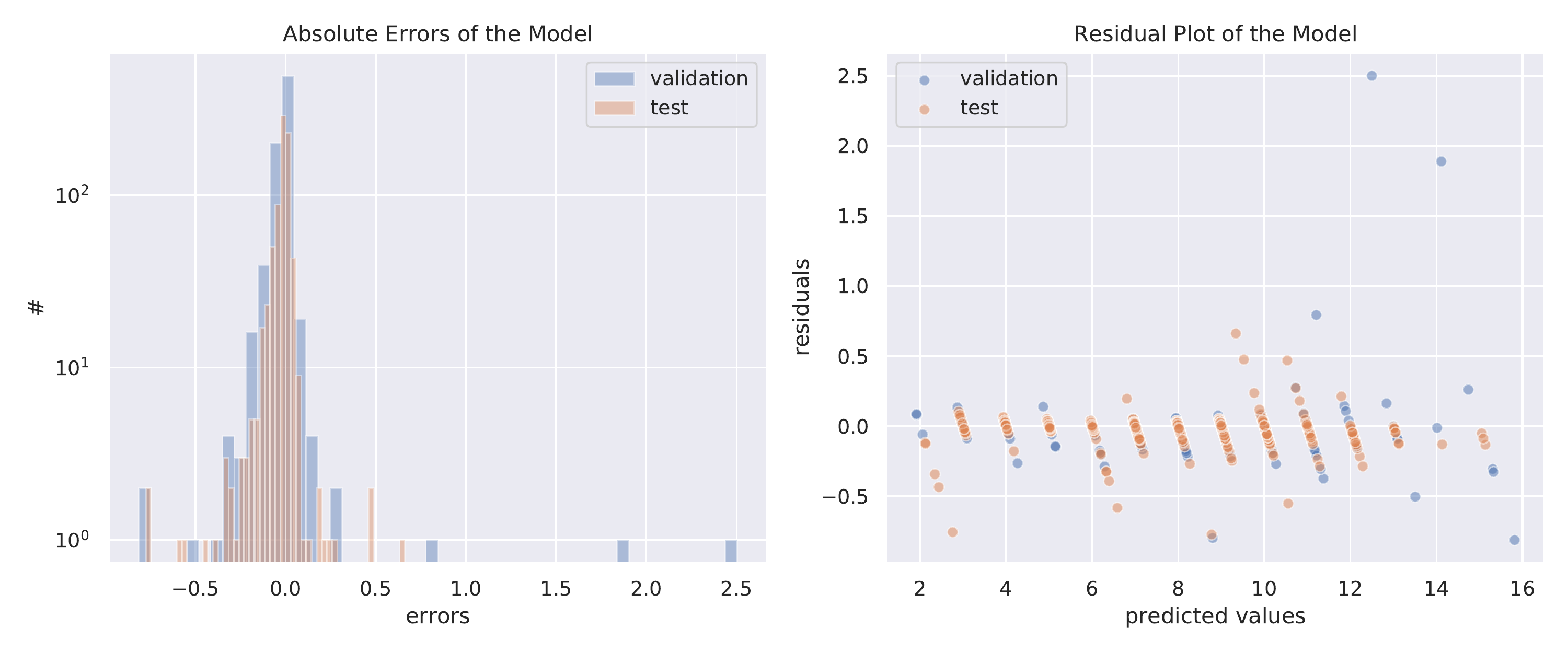}
\caption{%
  Difference between the true values of $h^{1,1}$ and their predictions seen as a univariate distribution (on the left) and as a function of the predicted value (on the right): the histogram shows the distribution and extension in values of the difference between true values and predictions, while the scatter plot exhibits the constant variance of the residual error (the network performs equally well over the entire range of $h^{1,1}$).
}
\label{fig:cnn:residuals}
\end{figure}

\begin{figure}[tbp]
\centering
\includegraphics[width=0.5\linewidth]{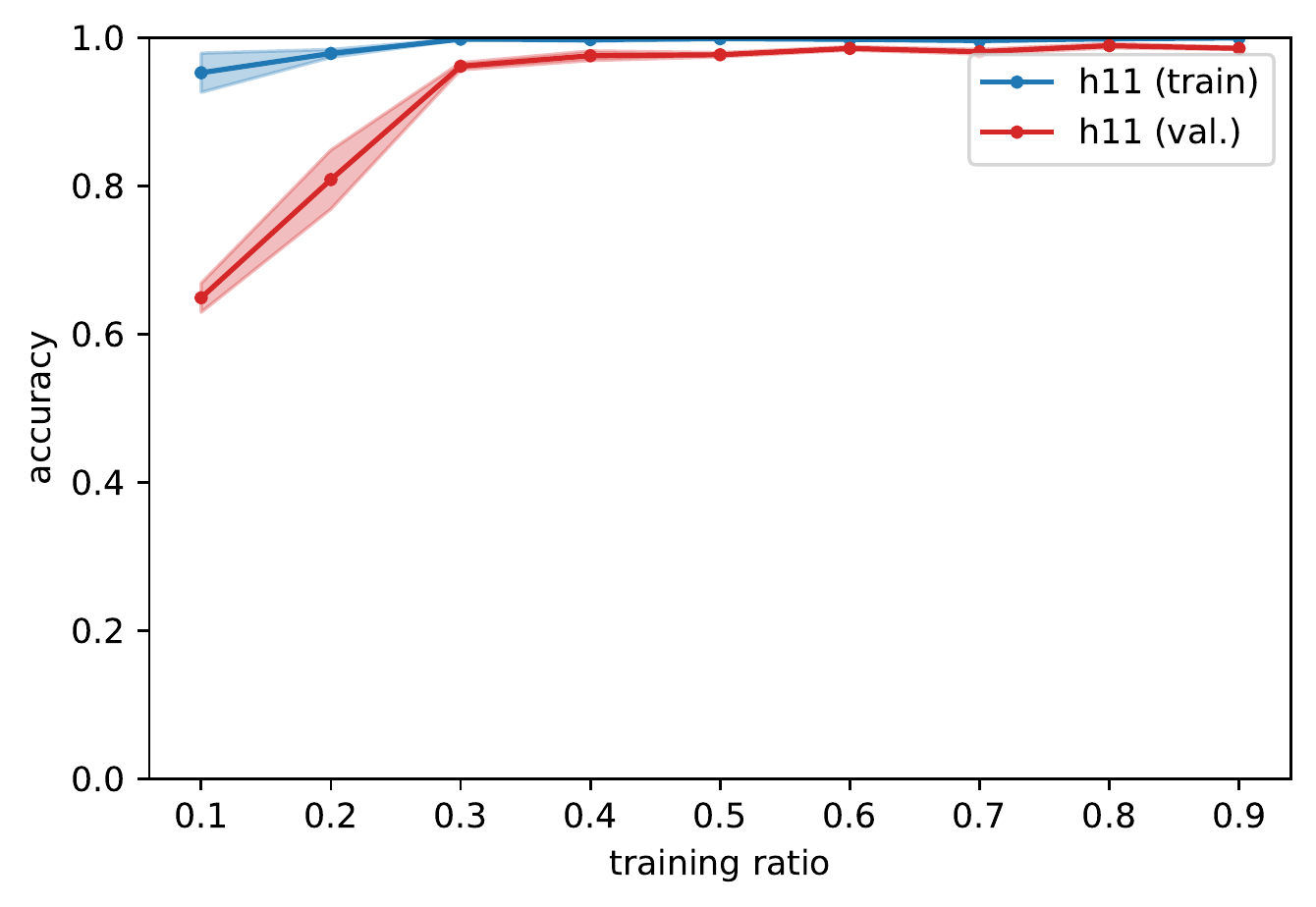}
\caption{%
  Learning curve of the Inception network.
  The colored area denotes the $1\sigma$ region.
}
\label{fig:cnn:lc}
\end{figure}

As presented in Figure~\ref{fig:cnn:residuals}, the network performs equally well over the entire range of $h^{1,1}$ both in the validation and test sets: the variance of the difference between the observed values of the Hodge number and its predictions (i.e.\ the residuals) is constant as shown by the scatter plot.
Moreover, the histogram of the residuals shows that the distribution is peaked around \num{0} and very few predictions lie far from the central value: the variance is in fact very small.

\subsection{Ablation study}
\label{sec:ablation}

\begin{figure}[tbp]
\centering
\includegraphics[width=\linewidth]{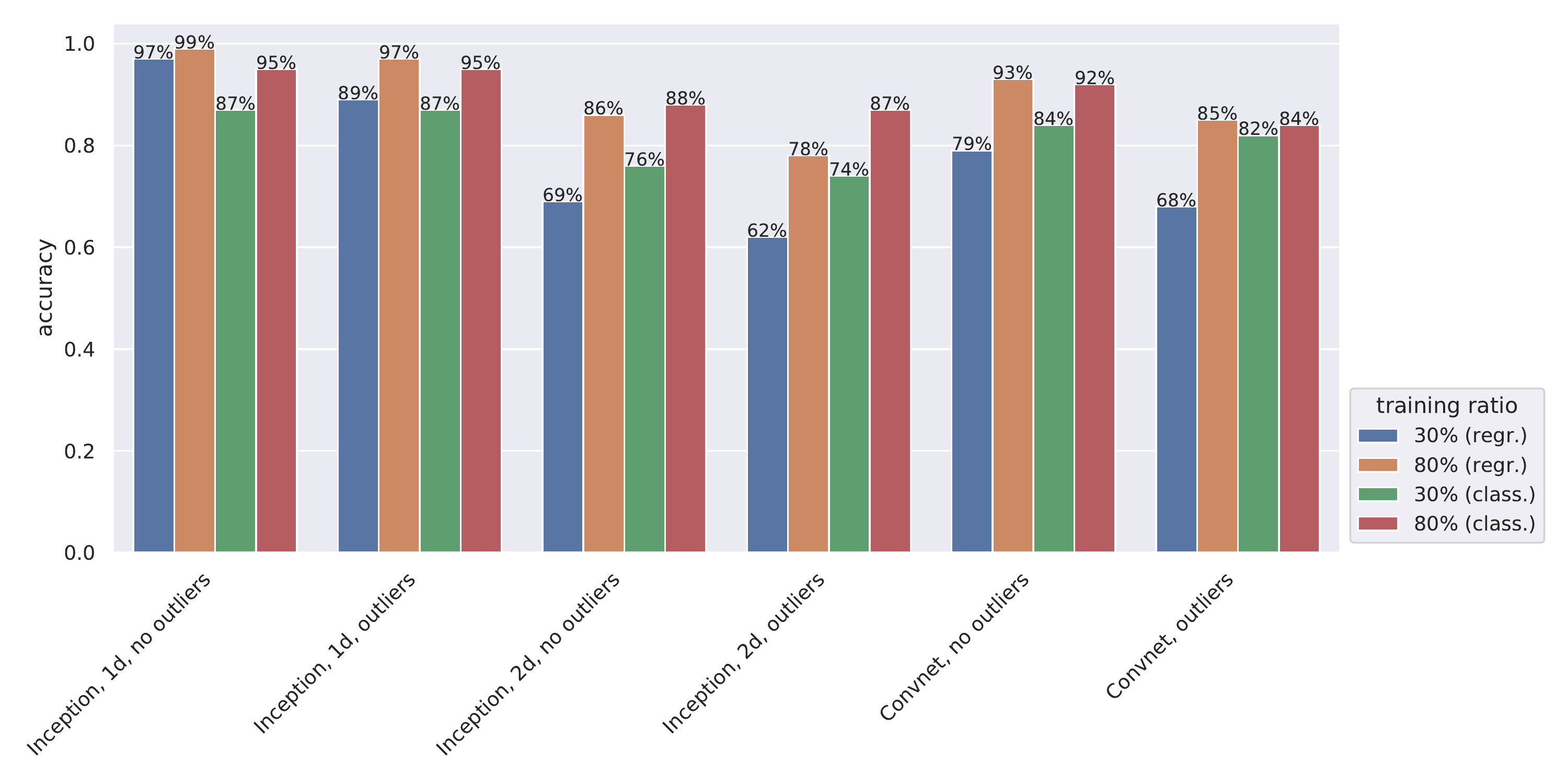}
\caption{%
  Comparison of accuracy for different properties.
  The label ``no outliers'' means that outliers are excluded from the training and validation set.
  The labels $1d$ and $2d$ refer respectively to the kernels $(12 \times 1, 1 \times 15)$ and $(3 \times 3, 5 \times 5)$.
}
\label{fig:cnn:ablation}
\end{figure}

We can now study in detail the relative impact of each improvement introduced in our paper.
The three points of comparison are 1) parallel vs sequential convolution layers, 2) using $1d$ kernels $12 \times 1$ and $1 \times 15$ or $2d$ kernels $3 \times 3$ and $5 \times 5$ (without changing the number of layers), 3) including or removing outliers from the training data.
A comparison of the accuracy achieved by different models is displayed in Figure~\ref{fig:cnn:ablation}.

First, we want to measure the benefit of using parallel instead of sequential convolutions.
In~\cite{Erbin:2020:MachineLearningMethodological} we have built a convolutional network (\emph{convnet} in Figure~\ref{fig:cnn:ablation}) made of $4$ layers with $180$, $100$, $40$ and $20$ units, all with a $5 \times 5$ kernel and $\ell_1$ and $\ell_2$ regularisation \num{e-4} and \num{e-3} (\num{\approx 580000} parameters).
The accuracies of this network at a few training ratios are given in Table~\ref{tab:cnn:inception_results} and we refer the reader to~\cite{Erbin:2020:MachineLearningMethodological} for more details.
While this network performs better than earlier models (compare Figures~\ref{fig:intro:compare} and~\ref{fig:cnn:ablation}), its accuracy is below the Inception model.

Second, we wish to uncover the effect of using $1d$ kernels $12 \times 1$ and $1 \times 15$ instead of $2d$ kernels.
For this, we have trained a new version of the Inception model with the $1d$ kernels replaced by concurrent $3 \times 3$ and $5 \times 5$ kernels (typical in computer vision tasks), leaving all other hyperparameters identical (\num{\approx 290000} parameters).
From Figure~\ref{fig:cnn:ablation}, we find that this network performs even less well than the sequential convolutional network.
One possible explanation is that the two $1d$ convolutional windows process separately the information of each single projective spaces (columns) or polynomial equation (rows), scanning all of them one after the other.
This could explain why it is necessary to have two $1d$ kernels: one for the projective spaces, one for the equations.

Third, we have argued that removing outliers from the training and validation sets helps the network to learn better.
The effect is not as important as the previous two points, but still noticeable (Figure~\ref{fig:cnn:ablation}).

Finally, we compared the difference between regression and classification.
We have one-hot encoded the Hodge numbers, replaced the last layer of the network described in subsection~\ref{sec:architecture} by a softmax, and used the cross-entropy loss for optimization.
We find that classification is less efficient than regression (Figure~\ref{fig:cnn:ablation}).
Adding two additional Inception modules brings the accuracy to $96\%$, still below the result from the regression network.

In conclusion, we see that convolutional layers working in parallel are responsible for a large part of the performance boost.
That convolution is useful for CICY may seem counter-intuitive~\cite{Bull:2018:CICY3Folds} since the configuration matrices are not rotation nor translation invariant but only permutation invariant.
However, we first note that convolution alone is only equivariant to global translation: it is not invariant to rotation nor translation (even locally), both of which require the addition of pooling layers (which we do not have)~\cite{Goodfellow:2016:DeepLearning}.
Moreover, convolution layers can be understood more generally as a way to spot different patterns in data by sharing weights, storing them in multiple channels, and recombining them in more complicated representations in subsequent layers.
For instance, the original Inception models~\cite{Szegedy:2014:InceptionNetwork, Szegedy:2015:RethinkingInceptionArchitecture, Szegedy:2017:Inceptionv4InceptionResNetImpact} include layers with $1 \times 1$ kernel, which clearly do not exploit invariance properties.
Another motivation for using convolution layers is parameter sharing: the same operations are applied at different locations of the input.
Parameter sharing with the $1d$ shape of the kernels implies that the same formulas are applied to each equation and each projective space, as can be expected for a geometric object.

\section{Conclusion}

We have introduced a new type of neural network to compute the Hodge number $h^{1,1}$ of complete intersection Calabi--Yau $3$-folds.
This neural network inspired by Google's Inception model gets near-perfect accuracy, using fewer data and parameters than existing models.
This improves largely the prediction power of the network and proves that deep learning is perfectly adapted for computations in algebraic topology.
Hence, this network should definitely be explored at length to exploit its potential, which seems to be as promising for theoretical physics and mathematics as it has been in computer vision.

The next step consists in predicting also the Hodge number $h^{2,1}$.
A preliminary analysis shows that the task is harder and the Inception network reaches only $\SI{50}{\percent}$ accuracy -- but it is higher than all other models, the best of which reach at most $\SI{35}{\percent}$ (for SVM with Gaussian kernel and sequential convolutional network)~\cite{Erbin:2020:MachineLearningMethodological}.
One solution is to use a better representation of the data.
A first possibility is to use the favourable representation from~\cite{Anderson:2017:Fibrations}, but this does not help~\cite{Erbin:2020:MachineLearningMethodological}.
Another more promising avenue is to use the graph representation introduced in~\cite{Krippendorf:2020:DetectingSymmetriesNeural}.
It will also be interesting to extend our analysis to other topological objects useful for string theory.
A last open question is to understand what the neural network learned and if it is possible to extract any interesting information from the weights.
We leave these questions for the future.

\section*{Acknowledgements}

We are grateful to Sven Krippendorf and Fabian Rühle for discussions, and to Siavash Golkar for comments on the draft.
The work of H.E.\ has been conducted under a Carl Friedrich von Siemens Research Fellowship of the Alexander von Humboldt Foundation for postdoctoral researchers during part of this project.
H.E.\ and R.F.\ are partially supported by the \textsc{Miur Prin} Contract \textsc{2015Mp2cx4} ``Non-perturbative Aspects of Gauge Theories and Strings''.

\section*{References}
\bibliographystyle{unsrt}
\bibliography{inception_network_cicy.bib}

\end{document}